\documentclass[12pt, draftcls, onecolumn]{IEEEtran}
\ifCLASSINFOpdf
  \usepackage[pdftex]{graphicx}
  \DeclareGraphicsExtensions{.pdf,.jpeg,.png}
\else
  \usepackage[dvips]{graphicx}
\fi
\usepackage{epstopdf}
\usepackage[cmex10]{amsmath}
\usepackage{amssymb}
\usepackage{wasysym}
\usepackage{algorithm}
\usepackage{algorithmic}
\usepackage{array}
\usepackage{cite}
\usepackage{color}
\usepackage{url}

\usepackage{epsfig,latexsym}
\usepackage{cite}
\usepackage{verbatim}
\usepackage{amsopn}
\usepackage{booktabs}
\usepackage{bm}
\usepackage{stfloats}
\usepackage{hyperref}
\usepackage{graphicx}
\usepackage{float}
\usepackage{subfigure}
\usepackage{makecell}

\begin{document}

\title{Hybrid Beamforming for Millimeter Wave MIMO Integrated Sensing and Communications}

\author{Chenhao~Qi,~\IEEEmembership{Senior~Member,~IEEE}, Wei~Ci,~\IEEEmembership{Student~Member,~IEEE}, \\
	Jinming~Zhang,~\IEEEmembership{Student~Member,~IEEE} and Xiaohu You,~\IEEEmembership{Fellow,~IEEE}
	\thanks{Chenhao~Qi, Wei~Ci, Jinming~Zhang and Xiaohu You are with the School of Information Science and Engineering, Southeast University, Nanjing 210096, China (Email: \{qch,ciwei,jmzhang,xhyu\}@seu.edu.cn).}
}

\markboth{}
{}

\maketitle

\begin{abstract}
In this letter, we consider hybrid beamforming for millimeter wave (mmWave) MIMO integrated sensing and communications (ISAC). We design the transmit beam of a dual-functional radar-communication (DFRC) base station (BS), aiming at approaching the objective radar beam pattern, subject to the constraints of the signal to interference-plus-noise ratio (SINR) of communication users and total transmission power of the DFRC BS. To provide additional degree of freedom for the beam design problem, we introduce a phase vector to the objective beam pattern and propose an alternating minimization method to iteratively optimize the transmit beam and the phase vector, which involves second-order cone programming and constrained least squared estimation, respectively. Then based on the designed transmit beam, we determine the analog beamformer and digital beamformer subject to the constant envelop constraint of phase shifter network in mmWave MIMO, still using the alternating minimization method. Simulation results show that under the same SINR constraint of communication users, larger antenna array can achieve better radar beam quality.

\end{abstract}


\section{Introduction}
As a candidate technology for the next generation wireless communications, integrated sensing and communications (ISAC) are attracting interests from academia, industry and government~\cite{LiuFanArxiv2021}. ISAC are capable of sufficiently sharing spatial, temporal, frequency and power resources as well as reducing hardware and software complexity for both wireless communications and radar sensing~\cite{AndrewZhangJSTSP2021,ZhengLeSPM2019}. Current study on signal processing for ISAC system can be mainly categorized into communication-centric design, sensing-centric design and joint design. On the other hand, the scarce of frequency resource for wireless communications motivates extensive exploration and application of millimeter wave (mmWave) frequency band~\cite{SciChinaChenhao2021}. In general, mmWave massive MIMO system uses phase shifter networks to form highly directional communication beams, which has a common feature with radar system using phased antenna array. Therefore, it is natural to give more focus on mmWave MIMO ISAC~\cite{MKJmmWaveSPM2019}.

One challenge for MIMO ISAC is the beamforming design. In~\cite{LiuFanMUMIMO2018}, the design of MIMO ISAC beamforming is investigated for separated deployment and shared deployment of radar and communication antennas, where the problem to design the beamformer to match an objective radar beam pattern subject to the constraints of communication performance is solved by semidefinite relaxation (SDR) optimization. In~\cite{multibeamTVT2019}, a multibeam framework is proposed for the time-division duplex (TDD) joint communications and radar, where the framework simultaneously uses fixed subbeam for wireless communications and packet-varying scanning subbeam for radar sensing  based on the same antenna array. In~\cite{LiuXiangJoint2020}, the transmit beamforming of MIMO dual-functional radar-communication (DFRC) system is studied, aiming at generating desired beam pattern and meanwhile decreasing cross correlation pattern at several given directions, subject to the power constraint for each transmit antenna and the signal to interference-plus-noise ratio (SINR) constraint for each communication user. Different from the work in~\cite{LiuFanMUMIMO2018,multibeamTVT2019,LiuXiangJoint2020}, in \cite{LiuFanICASSP2019} mmWave MIMO ISAC with hybrid beamforming are considered, where the design of hybrid beamformer is formulated as an optimization problem of the weight summation of the communication performance and radar performance subject to the constraints of constant modulus for analog beamformer and the total transmission power. It would be interesting to study the hybrid beamforming in mmWave MIMO ISAC system when further considering the SINR constraint for communication users. 


In this letter, for an mmWave MIMO ISAC system, we design the transmit beam of a DFRC base station (BS), aiming at approaching the objective radar beam pattern, subject to the SINR constraints of communication users and total transmission power of the DFRC BS. To provide additional degree of freedom for the beam design problem, we introduce a phase vector to the objective beam pattern and propose an alternating minimization method to iteratively optimize the transmit beam and the phase vector, which involves second-order cone programming (SOCP) and constrained least squared (LS) estimation, respectively. Then based on the designed transmit beam, we determine the analog beamformer and digital beamformer subject to the constant envelop constraint of phase shifter network in mmWave MIMO, still using alternating minimization method.

The notations are defined as follows. Symbols for matrices and vectors are denoted in boldface. $s, \bm{s}, \bm{S}$ denote a scalar, a vector and a matrix, respectively, while $(*)^{\rm T}, (*)^{\rm H}, \|\cdot\|_2, \|\cdot\|_{\rm F}$ denote the transpose, the conjugate transpose, the $\ell_2$ norm and the Frobenius norm, respectively. $\bm M(m,n)$ represents the entry located in the $m$th row and $n$th column of a matrix $\bm M$. $\bm{I}_K$ denotes a $K$-by-$K$ identity matrix. $\mathcal{CN}(m, \bm{R})$ represents the complex Gaussian distribution whose mean is $m$ and covariance matrix is $\bm{R}$. $\mathbb{C}$ and $\mathbb{R}$ represent the sets of complex-valued numbers and the real-valued numbers, respectively.


\section{System Model}\label{sec.system.model}
\begin{figure}[!t]
	\centerline{\includegraphics[height=8.5cm]{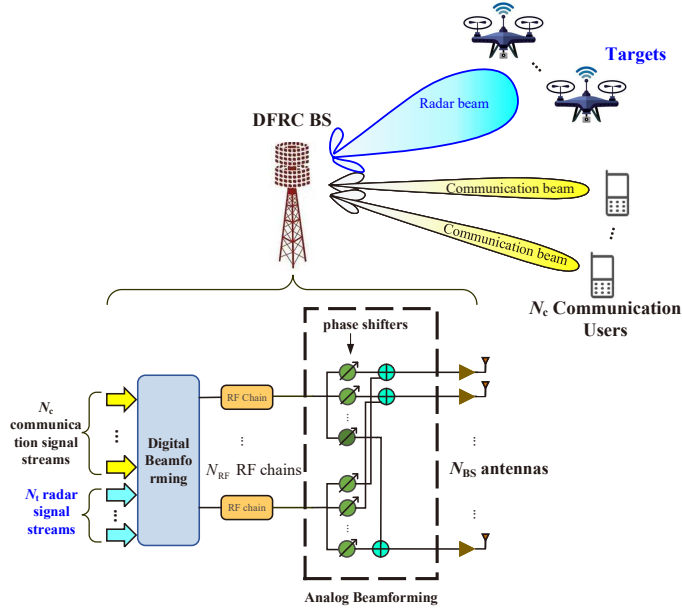}}
	\caption{Illustration of mmWave MIMO ISAC system.}
	\label{SystemModel}
\end{figure}

As shown in Fig.~\ref{SystemModel}, we consider an mmWave MIMO ISAC system, where a DFRC BS equipped with $N_{\rm BS}$ antennas serves $N_{\rm c}$ single-antenna communication users and the DFRC BS also wants to detect several targets. The $N_{\rm BS}$ antennas of the DFRC BS are placed in a uniform linear array (ULA) with half-wavelength interval, and work for both wireless communications and radar sensing. In fact, with a large $N_{\rm BS}$, it is not difficult to generate multi-mainlobe beam (also known as multibeam~\cite{multibeamTVT2019}), with each mainlobe pointing to a spatial region where the targets are possibly located~\cite{HierarchicalCodebookTWC2020}. 

To serve each communication user with an independent data stream, we need $N_{\rm c}$ radio frequency (RF) chains to generate $N_{\rm c}$ communication beams, with each user corresponding to a RF chain and a beam. To provide more degrees of freedom for target detection as well as giving dedicated RF chains for echo signal processing, we may need $N_{\rm t}$ RF chains for radar sensing. If $N_{\rm t}=0$ indicating there is no dedicated RF chains for radar sensing, the beamforming for target detection completely relies on the communication resource and the ISAC system design will be more challenging. Then the total number of RF chains used by the DFRC BS is
\begin{equation}\label{numberRFchain}
	N_{\rm RF} \triangleq N_{\rm c}+N_{\rm t}.
\end{equation} 

According to the Saleh-Valenzuela channel model widely used in mmWave MIMO wireless transmission, the channel between the DFRC BS and the $n$th communication user for $n=1,2,\ldots,N_{\rm c}$, can be expressed as a vector
\begin{equation}\label{widebandChannel}
	\bm {h}_n = \sqrt{\frac{N_ {\rm{BS}}}{L_n}}\sum_{l = 1}^{L_n} g_l^{(n)} \bm {\alpha}^{\rm H}(N_{\rm BS}, \theta_l^{(n)}),
\end{equation}
where $L_n$ is the number of resolvable channel paths. $g_l^{(n)}$ and $\theta_l^{(n)}$ denote respectively the channel gain and angle-of-departure (AoD) of the $l$th channel path, for $l=1,2,\ldots,L_n$. $\bm \alpha(N_{\rm BS}, \theta_l^{(n)})$ denotes the channel steering vector expressed as
\begin{equation}\label{ChannelSteeringVector1}
	\bm{\alpha}(N_{\rm BS}, {\theta_l^{(n)}})=\frac{1}{\sqrt{N_{\rm BS}}}[1,e^{j\pi\theta_l^{(n)}},\ldots,e^{j(N_{\rm BS}-1)\pi\theta_l^{(n)}}]^{\rm T}
\end{equation}
which is a function of $N_{\rm BS}$ and $\theta_l^{(n)}$.

For mmWave MIMO transmission, the hybrid beamforming architecture including analog beamforming and digital beamforming is typically adopted by the DFRC BS~\cite{SciChinaChenhao2021}. We denote the analog beamformer and digital beamformer as $\bm{F}_{\rm RF}$ and $\bm{F}_{\rm BB}$, respectively, where $\bm{F}_{\rm RF}\in \mathbb{C}^{N_{\rm BS} \times N_{\rm RF}}$ and $\bm{F}_{\rm BB} \in \mathbb{C}^{ N_{\rm RF} \times N_{\rm RF}}$. The transmitted signal before hybrid beamforming at the DFRC BS is denoted as $\bm{x}\in\mathbb{C}^{N_{\rm RF}}$, satisfying ${\rm E}\{\bm{x}\}=\bm{0}$ and ${\rm E}\{\bm{x} \bm{x}^{\rm H}\}=\bm{I}_{N_{\rm RF}}$. The front part of $\bm{x}$, including $x_1,x_2,\ldots,x_{N_{\rm c}}$, is the communication signal to be transmitted to the $N_{\rm c}$ communication users. The rear part of $\bm{x}$, including $x_{N_{\rm c}+1},\ldots,x_{N_{\rm RF}}$, is the radar waveform at some time instance. Note that in this work we only consider one time instance, with our focus on beamforming design in the spatial domain, while waveform design for multiple time instance in the time domain is skipped. Then the received signal by the $N_{\rm c}$ communication users, denoted as $\bm{y}\triangleq[y_1,y_2,\ldots,y_{N_{\rm c}}]^{\rm T}\in\mathbb{C}^{N_{\rm c}}$, can be expressed as
\begin{equation}
	\bm{y}=\bm{H} \bm{F}_{\rm RF} \bm{F}_{\rm BB}\bm{x} + \bm{\eta}
\end{equation}
where 
\begin{equation}
	\bm{H}\triangleq[\bm{h}_1^{\rm T},\bm{h}_2^{\rm T},\ldots,\bm{h}_{N_{\rm c}}^{\rm T}]^{\rm T}~\in\mathbb{C}^{N_{\rm c}\times N_{\rm BS}}
\end{equation}
is a channel matrix between the DFRC BS and the $N_{\rm c}$ communication users, and $\bm{\eta}$ is an additive white Gaussian noise (AWGN) vector satisfying $\bm{\eta}\sim\mathcal{CN}(\bm{0},\sigma^2\bm{I}_{N_{\rm c}})$.

To simplify the notation, we define
\begin{equation}\label{HybridPrecodingMatrix}
	\bm{F}\triangleq [\bm{f}_1,\bm{f}_2,\ldots,\bm{f}_{N_{\rm RF}}] \triangleq\bm{F}_{\rm RF} \bm{F}_{\rm BB}~\in\mathbb{C}^{N_{\rm BS}\times N_{\rm RF}}.
\end{equation}
where the $n$th column of $\bm{F}$ is denoted as $\bm{f}_n$, for $n=1,2,\ldots,N_{\rm c}$. Then the received SINR of the $n$th communication user, for $n=1,2,\ldots,N_{\rm c}$, can be expressed as
\begin{equation}
	\gamma_n = \frac{|\bm{h}_n \bm{f}_n|^2}{ \sum_{i=1,i \neq n}^{ N_{\rm RF} }|\bm{h}_n \bm{f}_i|^2 +\sigma^2}.
\end{equation}
Note that different communication user may have different requirement of wireless quality of service. For example, the video service requires higher SINR than the text service. By defining the threshold of SINR requirement of the $n$th user as $\Gamma_n$, we can write the SINR constraint as $\gamma_n \geq \Gamma_n$.

On the other hand, to detect multiple targets, the DFRC BS needs to generate various radar beams. As a simple example, the DFRC BS may scan different angle of space using the DFT codewords~\cite{TwoStepCodewordDesign2020}. Suppose the angle space of interest is sampled by $M$ points, $\phi_1$, $\phi_2$,\ldots,$\phi_M$, where $-1\leq\phi_1<\phi_2<\ldots<\phi_M\leq 1$. Larger $M$ results in finer sampling of the angle space and more precise description of the beam. Then the spatial sampling matrix based on $M$ sampling points can be denoted as
\begin{equation}\label{SamplingMatrix}
	\bm{\Phi}=\big[\bm{\alpha}(N_{\rm BS}, {\phi_1}),\bm{\alpha}(N_{\rm BS}, {\phi_2}),\ldots,\bm{\alpha}(N_{\rm BS}, {\phi_M})\big]^{\rm T}. 
\end{equation}
In fact, the transmit beam of the DFRC BS is $\sum_{i=1}^{N_{\rm RF}}\boldsymbol{f}_i$. Then the beam pattern of the transmit beam is $| \bm{\Phi}\sum_{i=1}^{N_{\rm RF}}\boldsymbol{f}_i |$, which is essentially to project the transmit beam on the channel steering vectors corresponding to the sampling points and then to obtain the absolute value of the projection. Note that the absolute value of a vector means obtaining the absolute value of each entry of the vector to form a same-dimensional vector.

Given an objective radar beam pattern $\bm{b}\in \mathbb{R}^{M}$, where each entry of $\bm{b}$ is nonnegative, we design the transmit beam of the DFRC BS, aiming at approaching $\bm{b}$, subject to the SINR constraint of communication users and the total transmission power of the DFRC BS. Then the transmit beam design problem can be formulated as
\begin{subequations}\label{hybrid precoding}\normalsize
	\begin{align}
		\underset{\bm{f}_1,\bm{f}_2,\ldots,\bm{f}_{N_{\rm RF}}}{\min} & \bigg\| \bm{D}\Big(\Big| \bm{\Phi}\sum_{i=1}^{N_{\rm RF}}\boldsymbol{f}_i \Big| - \bm{b} \Big)\bigg\|_2 \label{Objective}\\
		~~~\mathrm{s.t.} ~~~~~& \sum_{i=1}^{N_{\rm RF}} \| \bm{f}_i\|_2^2 \leq P_{\rm T} , \label{PowerConstraint}\\
		& \gamma_n \geq \Gamma_n,~n=1,2,\ldots,N_{\rm c}. \label{QoSConstraint}
	\end{align}
\end{subequations}
where $P_{\rm T}$ in \eqref{PowerConstraint} denotes the total transmission power of the DFRC BS and \eqref{QoSConstraint} is the SINR constraint of communication users. $\bm{D}$ is a predefined positive diagonal matrix with the diagonal entries being the weights at the corresponding sampling points of the angle space. Larger weight in $\bm{D}$ indicates higher requirement to approach $\bm{b}$ at the corresponding sampling points. 

\section{Hybrid Beamforming Design}
To provide additional degree of freedom for the transmit beam design in mmWave MIMO ISAC system~\cite{TwoStepCodewordDesign2020}, we introduce a phase vector $\bm{p}\in\mathbb{C}^M$ to the objective beam pattern $\bm{b}$, where $|\bm{p}_i| = 1$ for $i = 1,2,\cdots,M$. We further define a nonnegative diagonal matrix $\bm{A}\in\mathbb{R}^{M\times M}$, where the diagonal entries of $\bm{A}$ come from the corresponding entries of $\bm{b}$, i.e.,
\begin{equation}\label{DiagonalBeamGainMatrix}
	\bm{A}={\rm diag}\{\boldsymbol{b}\}.
\end{equation}
Then \eqref{hybrid precoding} can be rewritten as
\begin{subequations}\label{hybridPrecoding2}\normalsize
	\begin{align}
		\underset{\bm{f}_1,\bm{f}_2,\ldots,\bm{f}_{N_{\rm RF}},\bm{p}}{\min} & \bigg\| \bm{D}\Big( \bm{\Phi} \sum_{i=1}^{N_{\rm RF}}\boldsymbol{f}_i - \bm{A}\bm{p} \Big)\bigg\|_2 \label{Objective2}\\
		~~~\mathrm{s.t.} ~~~~~& \sum_{i=1}^{N_{\rm RF}} \| \bm{f}_i\|_2^2 \leq P_{\rm T} , \label{PowerConstraint2}\\
		& \gamma_n \geq \Gamma_n,~n=1,2,\ldots,N_{\rm c}. \label{QoSConstraint2}\\
		& |\bm{p}_i| = 1,~~i = 1,2,\cdots,M. \label{ConstantEnvelopConstraint2}
	\end{align}
\end{subequations}

To solve \eqref{hybridPrecoding2}, we propose an alternating minimization method, as follows.
\begin{enumerate}
\item Given $\bm{p}$, the optimization of $\bm{f}_1,\bm{f}_2,\ldots,\bm{f}_{N_{\rm RF}}$ in \eqref{hybridPrecoding2} can be expressed as
\begin{subequations}\label{hybrid precoding3}\normalsize
	\begin{align}
		\underset{\bm{f}_1,\bm{f}_2,\ldots,\bm{f}_{N_{\rm RF}}}{\min} & \bigg\| \bm{D}\Big( \bm{\Phi}\sum_{i=1}^{N_{\rm RF}}\boldsymbol{f}_i - \bm{A}\bm{p} \Big)\bigg\|_2 \label{Objective3}\\
		~~~\mathrm{s.t.} ~~~~~& \sum_{i=1}^{N_{\rm RF}} \| \bm{f}_i\|_2^2 \leq P_{\rm T} , \label{PowerConstraint3}\\
		& \gamma_n \geq \Gamma_n,~n=1,2,\ldots,N_{\rm c}. \label{QoSConstraint3}
	\end{align}
\end{subequations}

We define an auxiliary matrix
\begin{equation}\label{auxiliary matrix }
		\bm{S} \triangleq \big[\underbrace{\bm{I}_{N_{\rm BS}}, \bm{I}_{N_{\rm BS}},\cdots, \bm{I}_{N_{\rm BS}}}_{N_{\rm RF}}\big]~ \in\mathbb{R}^{N_{\rm BS} \times (N_{\rm RF}N_{\rm BS})}
\end{equation}
which essentially combines $N_{\rm RF}$ identity matrices $\bm{I}_{N_{\rm BS}}$ side by side. We further define
\begin{equation}
	\bm{f} \triangleq [\bm{f}_1^{\rm T},\bm{f}_2^{\rm T},\cdots,\bm{f}_{N_{\rm RF}}^{\rm T}]^{\rm T}~\in\mathbb{C}^{N_{\rm RF}N_{\rm BS}}
\end{equation}
which strings together different beamforming vectors.
Then the transmit beam of the DFRC BS can be rewritten as
\begin{equation}
	\sum_{i=1}^{N_{\rm RF}}\boldsymbol{f}_i= \bm{Sf}.
\end{equation}
We further define
\begin{equation}\label{auxiliary matrix }\normalsize
		\bm{S}_i \triangleq \big[\underbrace{\bm{0}, \cdots,\bm{0}}_{i-1},\bm{I}_{N_{\rm BS}},\underbrace{\bm{0},\cdots, \bm{0}}_{N_{\rm RF}-i}\big]~\in\mathbb{R}^{N_{\rm BS} \times (N_{\rm RF}N_{\rm BS})}
\end{equation}
for $ i = 1,2,\cdots,N_{\rm RF} $. Note that $\bm{S}_i$ is composed of $N_{\rm RF}-1$ zero matrices and an identity matrix $\bm{I}_{N_{\rm BS}}$. Then \eqref{QoSConstraint3} can be rewritten as
\begin{equation}\label{SINRconstraint2}
	\left\|\bm{t} \right\|_2 \leq \sqrt{1+\frac{1}{\Gamma_n}}\bm{h}_n\bm{S}_n\bm{f},~n = 1,2,\cdots,N_{\rm c}
\end{equation}
where
\begin{equation}
	\bm{t}\triangleq[\bm{h}_n\bm{S}_1\bm{f},~\bm{h}_n\bm{S}_2\bm{f},\cdots,
	\bm{h}_n\bm{S}_{N_{\rm RF}}\bm{f},~\sigma]^T~\in\mathbb{C}^{N_{\rm RF}+1}.
\end{equation}

Then \eqref{hybrid precoding3} can be rewritten as
\begin{subequations}\label{hybrid precoding4}\normalsize
	\begin{align}
		\underset{\bm{f}}{\min} & ~~\bigg\| \bm{D}\Big( \bm{\Phi}\bm{S}\bm{f} - \bm{A}\bm{p}\Big)\bigg\|_2 \label{Objective4}\\
		~~~\mathrm{s.t.} & ~~ \| \bm{f}\|_2^2 \leq P_{\rm T} ~{\rm and}~\eqref{SINRconstraint2}, \label{PowerConstraint4}
	\end{align}
\end{subequations}
which is a SOCP problem and can be solved using the existing optimization toolbox. 

\item Given $\bm{f}_1,\bm{f}_2,\ldots,\bm{f}_{N_{\rm RF}}$, the optimization of $\bm{p}$  in \eqref{hybridPrecoding2} can be expressed as
\begin{subequations}\label{hybrid precoding5}\normalsize
	\begin{align}
	\underset{\bm{p}}{\min} ~~ &\bigg\| \bm{D}\Big(\bm{\Phi}\sum_{i=1}^{N_{\rm RF}}\boldsymbol{f}_i - \bm{A}\bm{p}\Big)\bigg\|_2 \\
	~~~\mathrm{s.t.} ~~& |\bm{p}_i| = 1,~i = 1,2,\cdots,M  \label{phase-condition}
	\end{align}
\end{subequations}
which is a constrained LS estimation problem. Without the constraint of \eqref{phase-condition}, the unconstrained LS solution is 
\begin{equation}
	\bm{\widetilde{p}} = \bm{W}\sum_{i=1}^{N_{\rm RF}}\boldsymbol{f}_i
\end{equation}
where 
\begin{equation}
	\bm{W} \triangleq (\bm{A}^{\rm H} \bm{D}^{\rm H} \bm{DA})^{\rm -1} \bm{A}^{\rm H} \bm{D}^{\rm H} \bm{D}\bm{\Phi}
\end{equation}
is a constant matrix independent of optimization procedures and can be computed based on \eqref{SamplingMatrix} and \eqref{DiagonalBeamGainMatrix} before starting the optimization. Considering \eqref{phase-condition}, we denote the feasible solution to \eqref{hybrid precoding5} as $\bm{\widehat{p}}$, whose $i$th entry is
\begin{equation}
	\bm{\widehat{p}}_i = \frac{\bm{\widetilde{p}}_i}{|\bm{\widetilde{p}}_i|},~i=1,2,\ldots,M.
\end{equation}
		
\end{enumerate}

We alternatingly perform the above two steps 1) and 2) until a stop condition is triggered. The stop condition can be simply set that a predefined maximum number of iterations is reached. It can also be set that the objective function of \eqref{hybridPrecoding2} is smaller than a predefined threshold. 

\begin{algorithm}[!t]
	\caption{Hybrid Beamforming Design Scheme for mmWave MIMO ISAC system}
	\label{alg1}
	\begin{algorithmic}[1]
		\REQUIRE $\bm{H}$, $\bm{D}$, $\bm{\Phi}$, $\bm{b}$, $\sigma$, $\gamma_1$, $\gamma_2$, \ldots, $\gamma_{N_{\rm c}}$.
		\ENSURE $\bm{\widehat{F}}_{\rm RF}$, $\bm{\widehat{F}}_{\rm BB}$.
		\STATE Randomly generate $\bm{p}$ satisfying \eqref{ConstantEnvelopConstraint2}.
		\WHILE{\textit {Stop Condition 1} is not satisfied}
		\STATE Obtain $\bm{f}_1,\bm{f}_2,\ldots,\bm{f}_{N_{\rm RF}}$ by solving \eqref{hybrid precoding3}.
		\STATE Obtain $\bm{p}$ by solving \eqref{hybrid precoding5}.
		\ENDWHILE
		\STATE Randomly generate $\bm{F}_{\rm RF}$ satisfying \eqref{ConstantEnvelopConstraint}.
		\WHILE{\textit {Stop Condition 2} is not satisfied}
		\STATE Compute $\bm{{F}}_{\rm BB}$ via \eqref{LSestimationFBB}.
		\STATE Obtain $\bm{{F}}_{\rm RF}$ by solving \eqref{HybridBeamformingFRF}.
		\ENDWHILE
		\STATE Normalize $\bm{\widehat{F}}_{\rm BB}$ via \eqref{normalization}.
	\end{algorithmic}
\end{algorithm}

Suppose $\bm{\widehat{f}}_1,\bm{\widehat{f}}_2,\ldots,\bm{\widehat{f}}_{N_{\rm RF}}$ are obtained after finishing the above procedures. Similar to \eqref{HybridPrecodingMatrix}, we define 
\begin{equation}\label{HybridPrecodingMatrix2}
	\bm{\widehat{F}}\triangleq [\bm{\widehat{f}}_1,\bm{\widehat{f}}_2,\ldots,\bm{\widehat{f}}_{N_{\rm RF}}] ~\in\mathbb{C}^{N_{\rm BS}\times N_{\rm RF}}.
\end{equation}   

Based on the designed transmit beam of the DFRC BS, now we consider the hybrid beamformer design in terms of $\boldsymbol{F}_{\rm RF}$ and $\bm{F}_{\rm BB}$. Note that for mmWave MIMO wireless system, the analog beamformer is typically implemented by phase shifter networks, as shown in Fig.~\ref{SystemModel}. Therefore, we have constant envelop constraint for each entry of $\bm{F}_{\rm RF}$. Then the hybrid beamforming design problem to determine $\boldsymbol{F}_{\rm RF}$ and $\bm{F}_{\rm BB}$, given $\bm{\widehat{F}}$ in \eqref{HybridPrecodingMatrix2}, can be expressed as
\begin{subequations}\label{HybridBeamforming}
	\begin{align}
		\min_{\boldsymbol{F}_{\rm RF},\boldsymbol{F}_{\rm BB}} ~&\big\| \bm{\widehat{F}} -\boldsymbol{F}_{\rm RF}\boldsymbol{F}_{\rm BB} \big \|_{\rm F} \\
		\mathrm{s.t.} ~~~~& \big\| \boldsymbol{F}_{\rm RF}\boldsymbol{F}_{\rm BB} \big \|_{\rm F}^2 \leq P_{\rm T} \label{TotalPowerConstraint} \\
		& \big|\boldsymbol{F}_{\rm RF}(m,n)\big|=1,\label{ConstantEnvelopConstraint}\\
		&~m=1,2,\ldots,N_{\rm BS},~n=1,2,\ldots,N_{\rm RF} \nonumber
	\end{align}
\end{subequations}
where \eqref{ConstantEnvelopConstraint} is the constant envelop constraint due to the phase shifters and \eqref{TotalPowerConstraint} is the total transmission power constraint. In fact, we can temporarily neglect \eqref{TotalPowerConstraint} to solve \eqref{HybridBeamforming} and then normalize the obtained $\bm{F}_{\rm BB}$ to satisfy \eqref{TotalPowerConstraint}~\cite{yu2016alternating}. We still resort to the alternating minimization method and iteratively perform the following two steps.
\begin{enumerate}
	\item Given $\bm{F}_{\rm RF}$, the optimization of $\bm{F}_{\rm BB}$ in \eqref{HybridBeamforming} can be expressed as  
	\begin{equation}\label{HybridBeamformingFBB}
			\min_{\boldsymbol{F}_{\rm BB}} ~\big\| \bm{\widehat{F}} -\boldsymbol{F}_{\rm RF}\boldsymbol{F}_{\rm BB} \big \|_{\rm F}
	\end{equation}
	which is a LS problem with the solution as
	\begin{equation}\label{LSestimationFBB}
		\bm{\overline{F}}_{\rm BB} = (\bm{F}_{\rm RF}^{\rm H} \bm{F}_{\rm RF} )^{-1} \bm{F}_{\rm RF}^{\rm H} \bm{\widehat{F}}.
	\end{equation}

	\item Given $\bm{F}_{\rm BB}$, the optimization of $\bm{F}_{\rm RF}$ in \eqref{HybridBeamforming} can be expressed as 
	\begin{subequations}\label{HybridBeamformingFRF}
		\begin{align}
			\min_{\boldsymbol{F}_{\rm RF}} ~&\big\| \bm{\widehat{F}} -\boldsymbol{F}_{\rm RF}\boldsymbol{F}_{\rm BB} \big \|_{\rm F} \\
			\mathrm{s.t.} ~
			& \big|\boldsymbol{F}_{\rm RF}(m,n)\big|=1,\label{ConstantEnvelopConstraint3}\\
			&~m=1,2,\ldots,N_{\rm BS},~n=1,2,\ldots,N_{\rm RF} \nonumber
		\end{align}
	\end{subequations}
which is a typical Riemannian manifold optimization problem and can be solved by the existing toolbox.
\end{enumerate}

The above two steps 1) and 2) are iteratively performed until a stop condition is triggered. Supposing $\bm{\widehat{F}}_{\rm RF}$ and $\bm{\widehat{F}}_{\rm BB}$ are obtained, we finally normalize $\bm{\widehat{F}}_{\rm BB}$ to satisfy \eqref{TotalPowerConstraint} by 
\begin{equation}\label{normalization}
\bm{\widehat{F}}_{\rm BB}	\leftarrow \frac{ \sqrt{P_{\rm T}} }{ \| \bm{\widehat{F}}_{\rm RF} \bm{\widehat{F}}_{\rm BB} \|_{\rm F} } \bm{\widehat{F}}_{\rm BB}.
\end{equation}

The complete procedures for the hybrid beamforming design in mmWave MIMO ISAC system are summarized in \textbf{Algorithm~\ref{alg1}}.

\section{Simulation Results}\label{sec.simulation}
To evaluate the system performance, we consider a DFRC BS equipped with $N_{\rm BS}=128$ antennas and $N_{\rm RF}=3$ RF chains. Note that we set $N_{\rm t}=0$, indicating there is no dedicated RF chains to generate radar beam. The total transmission power of the DFRC BS is $P_{\rm T}=20{\rm dBm}$ and the AWGN noise power is $\sigma^2=0{\rm dBm}$. There are $N_{\rm c}=3$ communication users served by the DFRC BS, where each user has a line-of-sight (LoS) channel path and two non-line-of-sight (NLoS) channel paths between the user and the DFRC BS. The channel gain of LoS and NLoS obeys the distribution $g_1^{(n)}\sim\mathcal{CN}(0,1)$ and $g_2^{(n)},g_3^{(n)}\sim\mathcal{CN}(0,0.01)$ for $n=1,2,\ldots,N_{\rm c}$. The angle space of the mmWave MIMO ISAC system is $(-90^\circ,90^\circ]$ corresponding to the AoD space $(-1,1]$, which is equally sampled by $M=400$ points.
Three communication users are supposed to be located at $-70^\circ$, $-40^\circ$ and $-10^\circ$, i.e., $\theta_1^{(1)}=\sin(-70^\circ)$, $\theta_1^{(2)}=\sin(-40^\circ)$ and $\theta_1^{(3)}=\sin(-10^\circ)$. For simplicity, we set $\bm{D}=\bm{I}_M$. Suppose the objective beam pattern $\bm{b}$ has two bands, including one band $[10^\circ,30^\circ]$ and the other band $[40^\circ,60^\circ]$. The beam gain of $\bm{b}$ on the two bands is 
\begin{equation}
	\sqrt{ \frac{2 N_{\rm RF} P_{\rm T}}{\sin(30^\circ)-\sin(10^\circ)+\sin(60^\circ)-\sin(40^\circ)}}.
\end{equation}

To distinguish the curves in the figures, we name $\sum_{i=1}^{N_{\rm RF}}\boldsymbol{\widehat{f}}_i$ as the designed transmit beam (DTB) without hybrid beamforming (HBF), where $\bm{\widehat{f}}_1,\bm{\widehat{f}}_2,\ldots,\bm{\widehat{f}}_{N_{\rm RF}}$ is obtained after running steps 1 to 5 of \textbf{Algorithm~\ref{alg1}}. We define 
\begin{equation}
	\bm{\mathcal{F}} \triangleq \bm{\widehat{F}}_{\rm RF}\bm{\widehat{F}}_{\rm BB}
\end{equation}
where $\bm{\widehat{F}}_{\rm RF}$ and $\bm{\widehat{F}}_{\rm BB}$ are the output of \textbf{Algorithm~\ref{alg1}}. We name $\sum_{i=1}^{N_{\rm RF}}\boldsymbol{\mathcal{F}}_i$ as the DTB with HBF, where $\boldsymbol{\mathcal{F}}_i$ denotes the $i$th column of $\boldsymbol{\mathcal{F}}$ for $i=1,2,\ldots,N_{\rm RF}$.

\begin{figure}[!t]
	\centerline{\includegraphics[height=7.2 cm]{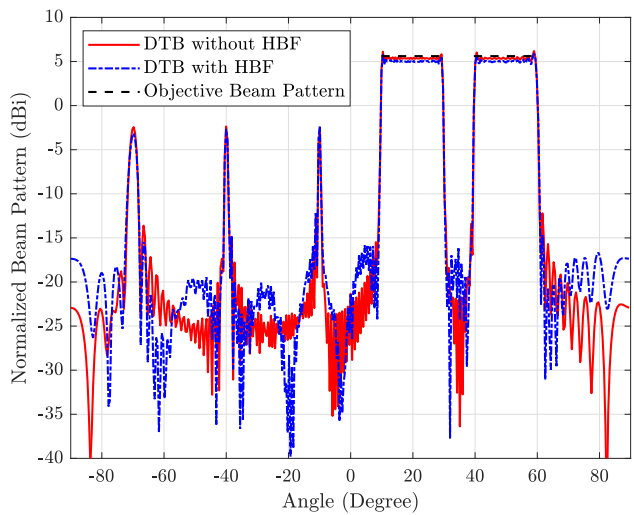}}
	\caption{Comparison of the beam pattern for the DTB with HBF and the DTB without HBF.}
	\label{BeamPattern}
\end{figure}

As shown in Fig.~\ref{BeamPattern}, we compare the beam pattern for the DTB with or without HBF, where the objective beam pattern $\bm{b}$ is provided as a performance bound. We normalize the beam pattern by a constant $\sqrt{N_{\rm RF} P_{\rm T}}$ to measure it in $\rm{dBi}$. We set $\Gamma_1=\Gamma_2=\Gamma_3=30{\rm dB}$. The three peaks in the figure correspond to the beams pointing at three communication users. Although the peaks are narrow, they occupy some energy, which causes the curves of the DTB with HBF or without HBF slightly lower than the performance bound within the two bands of $\bm{b}$. Due to the constant envelop constraint of phase shifter network in mmWave MIMO, there is energy leakage for the DTB with HBF, making its performance slightly worse than that of the DTB without HBF. 

\begin{figure}[!t]
	\centerline{\includegraphics[height=7.2 cm]{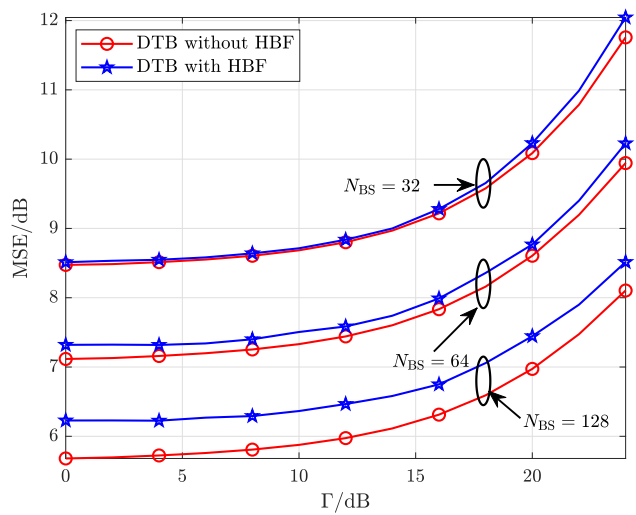}}
	\caption{Compare of the MSE of the beam pattern for the DTB with or without HBF in terms of different SINR constraint.}
	\label{MSE}
\end{figure}

We define the MSE of the beam pattern of the DTB without HBF as $\big\| |\bm{\Phi}\sum_{i=1}^{N_{\rm RF}}\boldsymbol{\widehat{f}}_i| - \bm{b} \big\|_2^2 / (N_{\rm RF} P_{\rm T})$ as a measurement of radar beam quality. The MSE of the beam pattern of the DTB with HBF can be similarly defined. As shown in Fig.~\ref{MSE}, we compare the MSE of the beam pattern for the DTB with or without HBF in terms of different SINR constraint of the communication users. It is seen that as $\Gamma$ increases, the MSE grows, which implies that higher requirement of communications results in worse radar beam quality. To analyze the impact of different number of antennas on the performance, we set $N_{\rm BS}=128, 64$ and $32$. It is seen that under the same SINR constraint of communication users, larger antenna array offers more degrees of freedom for the beam design and therefore can achieve better radar beam quality. We also observe that the MSE gap between the DTB with HBF and the DTB without HBF becomes larger, when $N_{\rm BS}$ increases. The reason is that more phase shifters of HBF result in more constraints in \eqref{ConstantEnvelopConstraint} and consequently larger errors between $\bm{\widehat{F}}$ and $\bm{\widehat{F}}_{\rm RF}\bm{\widehat{F}}_{\rm BB}$.  

\section{Conclusion}
In this letter, we have proposed an alternating minimization method to iteratively optimize the transmit beam and the phase vector. Then based on the designed transmit beam, we have determined the analog beamformer and digital beamformer subject to the constant envelop constraint of phase shifter network. Simulation results have shown that under the same SINR constraint of communication users, larger antenna array can achieve better radar beam quality. In the future, we will continue our work with the focus on performance optimization for mmWave MIMO ISAC.

\bibliographystyle{IEEEtran}
\bibliography{IEEEabrv,IEEEexample}
\end{document}